# A Framework for App Store Optimization

Artur Strzelecki[1[0000-0003-3487-0971]]

[1] University of Economics in Katowice, Katowice 40-287, Poland
artur.strzelecki@ue.katowice.pl

**Abstract.** In this paper a framework for app store optimization is proposed. The framework is based on two main areas: developer dependent elements and user dependent elements. Developer dependent elements are similar to factors in search engine optimization. User dependent elements are similar to activities in social media. The proposed framework is modelled after downloading sample data from two leading app stores: Google Play and Apple iTunes. Results show that developer dependent elements can be better optimized. Names and descriptions of mobile apps are not fully utilized.

## 1 Introduction

Together with increasing use of mobile devices there is increased supply of applications (apps) used on mobile devices. Software companies and developers create apps that installed on mobile device can be useful for mobile device owners. Now in two most popular app stores (iTunes and Google Play) are available millions of apps. Apps are offered in these stores in different categories of software, however still, competition is very high between apps in one category. Competition means that many apps offer similar functions and user can browse and choose from variety of selection.

This creates a need to propose a framework of app store optimization. App store optimization also can be described as app store marketing or mobile app search engine optimization. The focus is to improve ranking of mobile app directly within app store. Recently [1] noticed that developers make efforts to improve mobile app visibility in app store, but some of such actions can be treated as fraud. Especially when it comes to reviews and downloads, they can be pumped by developer. There is limited research on factors which are taken into account when apps ranking is created. Since apps are created by developers and used by users, ranking is influenced by these two groups. App store optimization has its roots in search engine optimization, where different factors are taken into account, when it comes to create ranking of websites. Earning from achievement in the field of search engine optimization author propose a framework for app store optimization. There are identified similarities between these two areas.

The research question is what factors are used for creating ranking of mobile apps directly in app store. Author in this paper propose a framework for ranking in app store optimization. This paper is organized as follows. Section 2 contains a review of the relevant literature on app stores and optimization in search engines area. In section 3 author describe the research method for choosing factors to propose framework. Section 4 contains characteristics of the sample date which are collected from iTunes and Google Play. In section 5 author highlight the contribution of the research and suggest possible implications of results, analyze current limitations of the research, draw conclusions and present ideas as for their future research on app store optimization.

## 2 Literature review

In literature review two areas are explored. First is about efforts made by app developers to promote apps in app store and user's which rate and review apps. Developer creates app and deliver it to app store. After software is delivered it can be offered with free app offers, continuous quality updates, investment in less popular (less competitive) categories and price changes [2]. New releases are found to change user opinion on app [3]. Number of rating, number of reviews, number of downloads are always positively changed, since they only can rise. Some developers update their apps very frequently, even once a week or twice a month. They are not too concerned about detailing the content of new updates and users are not too concerned about such information, whereas users highly rank frequently-updated apps instead of being annoyed about the high update frequency [4]. App rating is assigned to an app over its entire life time is aggregated into one rating that is displayed in the app store. However many apps do increase their version-to-version rating, while the store-rating of an app is resilient to fluctuations once an app has gathered a substantial number of raters [5]. The approach is proposed to assisting developers to select the proper release opportunity based on the purpose of the update and current condition of the app [6]. Developer also can do a shady moves to fraud app ranking. [1] defined two methods of fraud: inflate the app number downloads and ratings in a very short time.

User can publish review about app. Reviews have a major influence on the user's purchase decision [7]. Average rating according to the star principle as well as the number of reviews given determine the buying decision of an app to a very large degree. Review area is explored when it comes to see what the sentiment of the review is. Review can have positive or negative sentiment [8]. Analyzing reviews is done by text mining [9]. Most of the feedback is provided shortly after new releases, with a quickly decreasing frequency over time. Reviews typically contain multiple topics, such as user experience, bug reports, and feature requests [10]. Reviews are source for users' feedback, requests for new features or reporting bugs. Reviews represent feature requests, i.e. comments through which users either suggest new features for an app or express preferences for the re-design of already existing features of an app [11]. Reviews can be specific, as specific is an app. Analyzing feedbacks from a health and fitness-tracking app shows that the users of health and fitness-related apps are concerned about their physical activity records and physiological records. The records include track, distance, time, and calories burned during jogging or walking. App store reviews are used to analyze different aspects of app development and evolution [12]. There are proposed frameworks to acquire reviews in large number, extract informative user reviews by filtering noisy and irrelevant ones, then group the informative reviews automatically using topic modeling [13, 14]. There also systematic literature reviews where of opinion mining studies from mobile app store user reviews, which describes and compares the areas of research that have been explored thus far, drawing out common aspects, in app store analysis [15, 16].

Second area in literature review is search engine optimization. Since app store ranking is generated based on different factors, drawing from area of web search engines can help to build framework for app store optimization. Web search engines use different factors identified on and off website to determine ranking of certain webpage in search engines. In the beginning ranking factors were limited only to few elements taken into account when search engine results pages were created [17, 18]. Nowadays this topic is more explored and can be divided into onsite factors and offsite factors. Onsite factors are domain-related, website-related and page related [19]. Offsite factors are link-related [20], user-action-related [21], special-rules-related [22], brand-related [23] and spam-related [24].

## 3 Framework model

App stores distribute apps through the app store and have additional information about the app. A set of information is initially set by the developer. The app is delivered by the app developer. The developer sets a name for the app and creates a description of app features. The app is published with the new original url. The developer sets genre content rating and define system requirements for an app. Apps can be distributed through the app store for free or user needs to pay to download. Free apps can also offer in-app purchases for additional features. Another set of information is created after app is released. Users are downloading the app and make the number of downloads growing. User can also create reviews and rate an app in range of scale from 1 to 5.

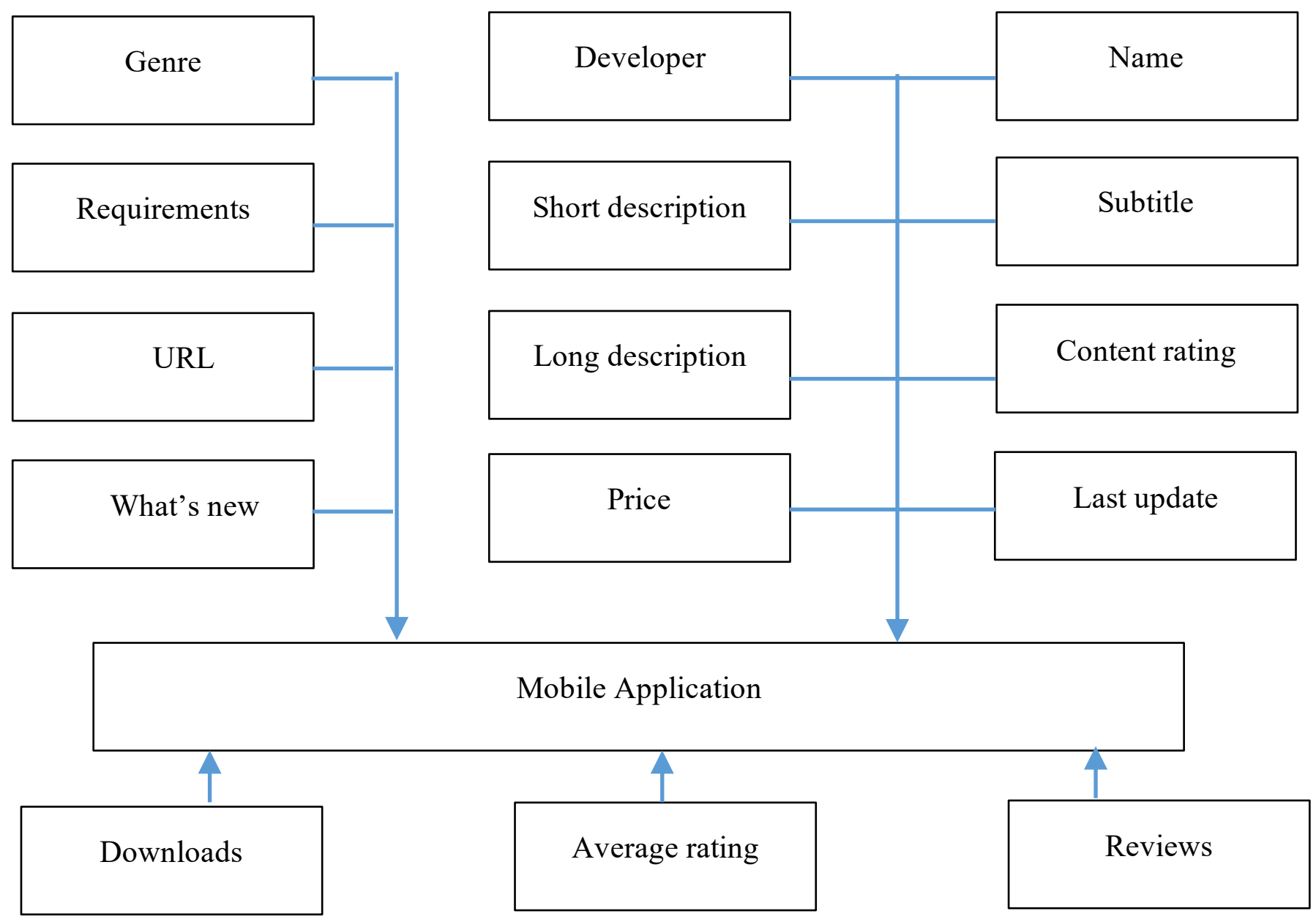

**Fig. 1.** A framework for app store optimization

Data for analysis and further usage in framework for app store optimization were collected from Google Play and Apple iTunes. Author used an automated software for websites crawling - Screaming Frog SEO Spider, which retrieved data, divided into several groups based on their type. Two different techniques for retrieving were used. First technique applied regular expressions to match elements like number of downloads, last date of update, content rating, range of pricing and software requirements. The second technique used CSSPath to match the next elements like developer, number of reviews, category, name, average rating. The URL was also retrieved for each application during the crawling process.

These two different techniques were needed, because some of the elements are written down firmly into a website structure and are always placed in the same context. These were retrieved by the CSSPath. The rest of elements can change their position in website structure, due to the incomplete data provided by the developer. Some of the apps in Google Play and Apple iTunes do not have all of the information usually displayed on app store. Regular expressions helped to collect the published in different part of websites.

Framework proposition is based on two areas. One area depends on the developer. Its content and settings are provided when the app is initially released in app store. Second area depends on users. If the app is being popular among user, they are starting to download and create reviews or rate the app in the app store.

### 3.1 Framework elements dependent on developer

**Developer.** The name of the developer is also a ranking factor, affecting directly the position of the application itself. The positive history of the developer affects the better evaluation of the application in the search ranking. Keywords in the developer's name affects each of its applications.

**Name.** The name of the application is important both for the app store optimization and for the experience. Often observed solution made by developers is to create a name by combining the brand name and the most important key words for the app. Google Play limits the name to 50 characters, whereas Apple iTunes limits the name to 30 characters.

**Subtitle.** (Only in Apple iTunes) Subtitle is placed right below the title and brings additional information on mobile app. It complements the app name by communicating the purpose and value of app in details and is limited to 30 characters.

**Genre.** Genre is a category for software. Currently Google Play offers 31 categories for mobile apps and Apple iTunes offers 28 categories.

**Description.** App stores allows to prepare two types of description, short and long version. The short version is only visible in mobile app store and its maximum length is 80 characters for Google Play and 170 characters for Apple iTunes. Long version is also visible in a desktop version and the maximum size is 4000 characters. However, only around 250 characters is visible after description is displayed, the rest is hidden. It can be showed after clicking button “Read more”.

**Content rating.** A content rating rates the suitability of mobile applications for its audience. It tells what age group is suitable to use mobile app.
**Requirements.** App store providers require that new apps target operating system version. System version is set in requirements. For new apps requirement in Google Play is at least Android 8.0 and in Apple App Store is at least iOS 12. Older apps can have lower operating system requirements.
**URL.** The URL can only be defined when the application is published, hence it is an element worth refining.
**Last update.** (Only in Google Play) For the application store, security is important, therefore the factor that has a ranking importance, but also affects the opinions about the application, is the frequency of its updating. It is necessary to update the application not only at the time of major changes to the functionality of the app, but also at every subsequent update of the operating system.
**What's new.** This element is intended to describe the updates introduced into the app. Developers can change the contents of the section only after the new app version is submitted to the store.

### 3.2 Framework elements dependent on users.

**Downloads.** (Only in Google Play) Downloads is the number of app installation on devices worldwide. Google Play publishes the only threshold number that has been reached.
**Average rating.** Average rating is a number with one decimal place in the scale from 1 to 5 and it is an aggregated rating from all ratings given by users.
**Reviews.** Apart from ratings users can write a review about an app. Reviews are a source for users' feedback, requests for new features or reporting bugs [25].

## 4 Data and Results

### 4.1 Data

In the process of app store mining, author has downloaded from Google Play about 50 000 apps, which belong to 31 different categories. Additionally, there is the main category Games which includes 17 subcategories (action, adventure, arcade, board, card, casino, casual, educational, music, puzzle, racing, role playing, simulation, sports, strategy, trivia, word). In the downloaded sample games take 16,66% of total number of applications, and other apps belong to 31 different categories. Figure 2 represents shares of all categories in this sample.

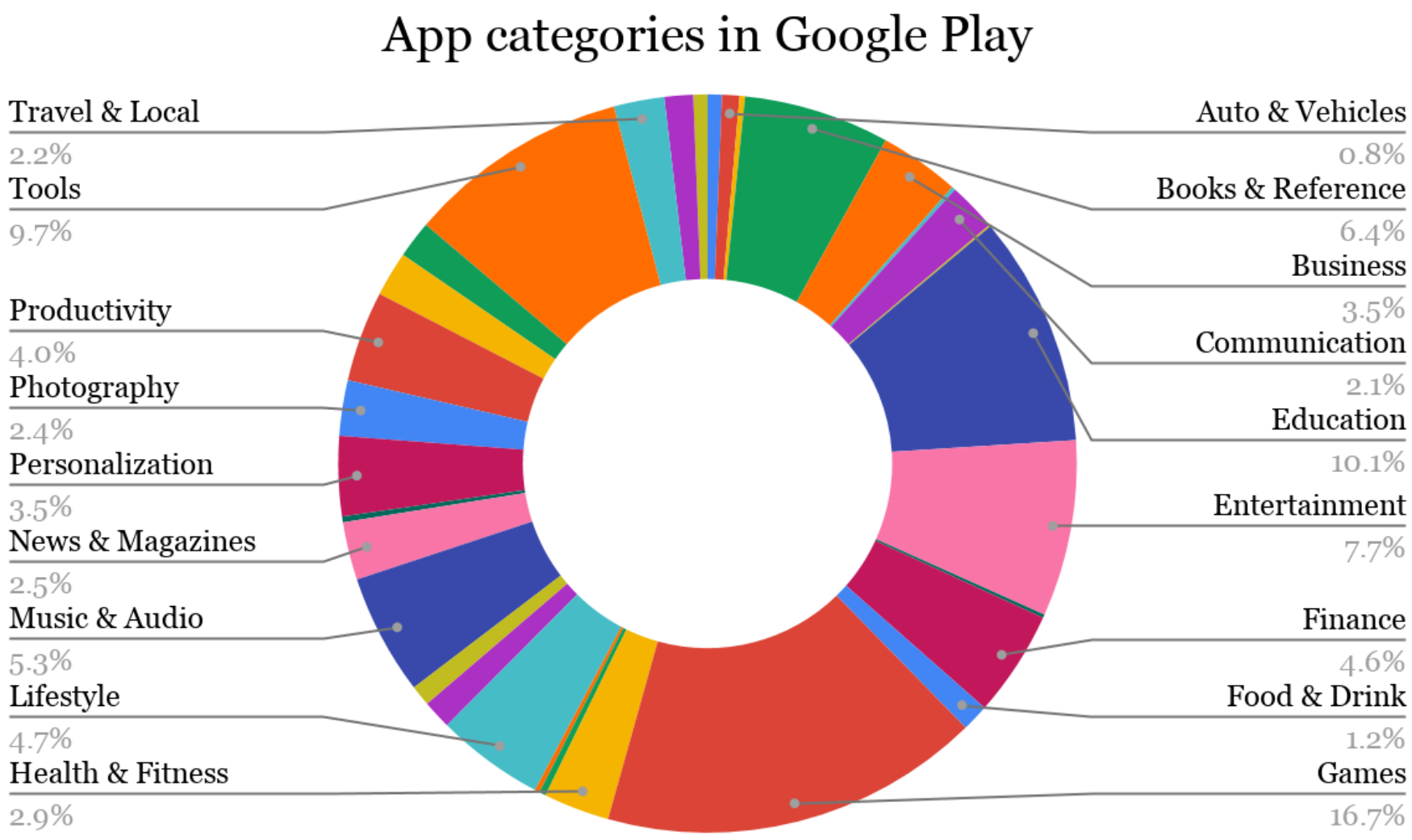

**Fig. 2.** App categories in Google Play store

Table 1 shows the descriptive statistics of 49 990 sample apps downloaded from Google Play and 6040 sample apps downloaded from Apple iTunes. Based on downloaded data it contains main characteristics. Table 1 also contains framework elements dependent on developers based on Google Play and Apple iTunes.

**Table 1.** Descriptive statistics of samples from Google Play and Apple iTunes

| App Store | Google Play | Apple iTunes |
|---|---|---|
| Number of downloaded apps | 49990 | 6040 |
| Number of developers | 16912 | 2786 |
| Number of genres | 31 | 25 |
| Avg length of name | 23 | 23 |
| Avg length of subtitle | na | 26 |
| Avg length of description | 1338 | 1694 |
| Median length of description | 1052 | 1489 |
| Apps with reviews | 97% | 79% |
| Average number of avg rating | 4,29 | 4,35 |

Autor downloaded smaller sample from Apple iTunes because this app store prevents against the massive download of data and after a few hundreds of downloaded apps, it turns on the HTTP response code 403 on every page with an app. Table 2 contains framework elements dependent on users based on Google Play like number of downloads, average number of reviews in download threshold and average from average rating in download threshold.

**Table 2.** Framework elements dependent on users.

| Download threshold | Number of apps | Avg of reviews | Avg from avg rating |
|---|---|---|---|
| 0 | 67 | 21 | 4,35 |
| 1 | 217 | 2 | 4,90 |
| 5 | 158 | 61 | 4,76 |
| 10 | 767 | 4 | 4,69 |
| 50 | 542 | 4 | 4,57 |
| 100 | 2324 | 8 | 4,43 |
| 500 | 1517 | 15 | 4,36 |
| 1000 | 5025 | 37 | 4,25 |
| 5000 | 2933 | 86 | 4,22 |
| 10000 | 8502 | 283 | 4,26 |
| 50000 | 4392 | 785 | 4,27 |
| 100000 | 10016 | 2757 | 4,27 |
| 500000 | 3753 | 9020 | 4,29 |
| 1000000 | 5833 | 31539 | 4,31 |
| 5000000 | 1520 | 102576 | 4,32 |
| 10000000 | 1813 | 343902 | 4,34 |
| 50000000 | 322 | 1113916 | 4,37 |
| 100000000 | 244 | 3541721 | 4,39 |
| 500000000 | 27 | 9739412 | 4,41 |
| 1000000000 | 17 | 26792154 | 4,27 |
| 5000000000 | 1 | 39797335 | 4,40 |

### 4.2 Results

Data in the Google Play store adapts to language settings of user's browser and operating system. During initial screening the authors have seen different settings coming from different languages. First, prices of apps itself or in-app purchases were displayed in the currency set in the web browser. Second, the types of content rating were different for different localization settings. Author checked three options of language & country setting: Poland, Ukraine and US. For Poland and Ukraine, the rating is displayed in Pan European Game Information (PEGI) standard, while for US the rating format comes from the Entertainment Software Rating Board (ESRB) [26]. Third, except for recommending applications on the basis of language and country, Google Play provides different search results for authorized and non-authorized users. If not logged in into Google account, a user is suggested a set of applications that differs from the one an authorized user will get (although the query in both situations is exactly the same.

Framework elements depending on developers reveal, that they do not use fully all of framework elements. Name of the app can be maximum 30 characters long, however average name length is 23 characters. This suggest, that there is still space that can be used by developers in name element. Subtitle is an optional element in Apple iTunes.

Not every app has set subtitle. In sample 41% of apps did not have subtitle. Subtitle can be maximum 30 characters long and the average length is 26 characters. This suggest, that if the subtitle is used, developers use it in its maximum capacity. Long description can be maximum 4000 characters long, however average length is 1338 characters and the median 1056 characters long for Google Play and average 1755 characters and median 1534 characters for Apple iTunes. This suggest, that there is still space that can be used by developers in description element.

Framework elements depending on users reveal that the larger download threshold is achieved, the more average reviews apps have. It is linear relationship, which is expected, since more downloads can result in more reviews. However, the average rating for apps with lowest download threshold is the highest and then for next thresholds is decreasing. Average rating is decreasing to 4,22 with threshold of 5000 downloads and then is increasing to threshold of 500 million downloads. Users' framework is built on data divided according to download threshold. It shows, that if app is more and more downloaded, the overall rating from user is growing. Usually developers are taking into account requests made in reviews and update apps with new and requested features.

## 5 Conclusion and discussion

In this paper a framework for app store optimization is proposed. The framework is based on two areas. One area is dependent on developer. The developer sets the initial setting for each app, which is distributed to the app store. After the app is being distributed, users' engagement in app distribution is second area. Users create number of downloads, reviews and overall rating. A framework is proposed for both leading app stores: Google Play and Apple iTunes.

Results show that developers are not fully using elements which are dependent on them. The lowest use is with description, in Google Play, developers on average use 33% of its capacity, and in Apple iTunes developers on average us 41% of its capacity. Second element which still has some space to use is the name of an app. In both app stores, on average developers use 76% of its full capacity.

This paper is a first attempt to create framework, which will explain, what elements are taken into account, when the ranking is created in app stores. Framework is created on data downloaded from two leading app stores: Google Play and Apple iTunes. This framework reveals that this ranking is depending on more than dozen factors. Some of these factors are identified in this paper. Author divided them into two groups, where one is depending on app developers and the second is depending on the users' engagement.

Proposed framework has some limitations. First is that the framework is proposed only on data that is publicly visible and accessible. Perhaps there are some other elements taken into this framework, not visible for users. This could be number of app uninstalls from device, number of app removals from the app store. Second, it takes only data from two app store. There are other app stores like Windows Phone or BlackBerry World, which were not taken into building this framework. This could future direction of research, to take also data from these stores and enhance proposed framework.